\documentclass{PoS}

\usepackage{amsmath}
\usepackage{soul}
\usepackage{cite}

\newcommand{\be}{\begin{equation}}
\newcommand{\ee}{\end{equation}}
\newcommand{\bea}{\begin{eqnarray}}
\newcommand{\eea}{\end{eqnarray}}

\newcommand{\nn}{\nonumber}
\newcommand{\Tr}{\textrm{Tr}}
\newcommand{\xv}{{\mathbf x}}
\newcommand{\vecnul}{{\mathbf 0}}
\newcommand{\bra}{\langle}
\newcommand{\ket}{\rangle}

\newcommand{\eps}{\epsilon}

\newcommand{\cC}{${\cal C}\;$}

\newcommand{\half}{\frac{1}{2}}

\def\ltap{\;\raisebox{-.5ex}{\rlap{$\sim$}} \raisebox{.5ex}{$<$}\;}
\def\gtap{\;\raisebox{-.5ex}{\rlap{$\sim$}} \raisebox{.5ex}{$>$}\;}

\title{Probing parity doubling in nucleons at high temperature}

\ShortTitle{Nucleons and parity doubling}

\author{Gert Aarts$^a$,
\speaker{Chris Allton}$^{a,b \dagger}$,
Simon Hands$^a$,
Benjamin J\"ager$^a$,
Chrisanthi Praki$^a$,
Jon-Ivar Skullerud$^{c,d}$
\\

$^a$Department of Physics, College of Science, Swansea University,
Swansea SA2 8PP, United Kingdom\\

$^b$All Souls College, Oxford OX1 4AL, United Kingdom\\

$^c$Department of Mathematical Physics, National University of Ireland
Maynooth, Maynooth, County Kildare, Ireland\\

$^d$School of Mathematics, Trinity College, Dublin 2, Ireland\\

$\dagger$ E-mail: c.allton@swan.ac.uk
}

\abstract{ The spectrum of nucleons and their parity partners is
  studied as a function of temperature spanning the deconfinement
  transition. We analyse our results using the correlation functions
  directly, exponential fits in the hadronic phase, and the Maximum
  Entropy Method. These techniques all indicate that there is
  degeneracy in the parity partners' channels in the deconfined
  phase. This is in accordance with the expectation that there is
  parity doubling and chiral symmetry in the deconfined phase. In the
  hadronic phase, we also find that the nucleon ground state is
  largely independent of temperature, whereas there are substantial
  temperature effects in the negative parity channel.  All results are
  obtained using our {\sc fastsum} 2+1 flavour ensembles.}

\FullConference{The 33rd International Symposium on Lattice Field Theory\\
		14 -18 July 2015\\
		Kobe International Conference Center, Kobe, Japan*}

\begin{document}



\section{Introduction}

Chiral symmetry in the mesonic sector has been extensively studied at
non-zero temperature \cite{Rapp:1999ej} but, due to the extra level of
difficulty, there have been few such studies in the baryonic sector.
On the lattice, baryonic analyses have been confined to a quenched
study of parity doubling \cite{Datta:2012fz} and screening mass
studies at both zero \cite{DeTar:1987ar,Datta:2012fz} and non-zero
density \cite{Pushkina:2004wa}.  The theoretical expectation is that
when chiral symmetry is restored (as occurs above the deconfining
temperature, $T_c$) parity partners in the baryon sector become
degenerate.

In nature, at zero temperature (where chiral symmetry is broken) the
positive parity nucleon state has mass $M_N= 939$ MeV whereas its
negative parity partner has a mass of $M_{N\ast}=1535$ MeV.  The
status of the parity channels above $T_c$ is not yet clear from
experiment, but since protons are routinely measured in heavy-ion
collisions, it is of great importance to study the pattern of parity
doubling theoretically.

In this talk we study full QCD lattice nucleon correlators and their
parity properties using a variety of techniques for temperatures
in both hadronic and plasma phases.
Full results appear in \cite{Aarts:2015mma}, but we extend this work
by including new Maximum Entropy Method (MEM) analyses \cite{future}.



\section{Nucleon Correlation Functions}

We use one of the standard interpolation operators for the nucleon, i.e.
\be
O_N(\xv,\tau) = \eps_{abc} u_a(\xv,\tau) \left[ u_b^T(\xv,\tau)
  {\cal C} \gamma_5 d_c(\xv,\tau) \right],
\nn
\ee
where $u, d$ are the
quark fields, $a,b,c$ are colour indices, other indices are suppressed
and \cC denotes the charge conjugation matrix.
We can define positive and negative parity operators and their
corresponding zero momentum correlators as
\be
\nn
O_{N_\pm}(\xv,\tau) = P_\pm O_N(\xv,\tau),
\quad\quad
G_\pm(\tau) = \int d^3x\, 
\left\bra \Tr\; O_{N_\pm}(\xv,\tau) \overline{O}_{N_\pm}(\vecnul,0) \right\ket,
\ee
where the parity projector is $P_\pm=\half(1\pm\gamma_4)$ and the
trace indicates the sum over Dirac indices, see \cite{Aarts:2015mma}.
For systems with chiral symmetry, a chiral rotation on the quark
fields shows that the two parity channels are degenerate, and,
up to overall minus signs \cite{Gattringer:2010zz},
\be
\nn
 G_\pm(\tau) = G_\mp(\tau) = G_\pm(1/T-\tau),
\ee
where the temporal extent of the lattice is the inverse temperature,
$N_\tau a_\tau = 1/T$.
Using time reflection properties, $G_\pm(\tau)$ contains
both parity channels each propagating either forwards or backwards
in time.
For this reason we consider $G_+(\tau)$ only and denote this as
$G(\tau)$ from now on.



\section{Lattice details} 


\begin{table}[t]
\begin{center}
\begin{tabular}{ccccccc}
    \hline
      $N_s$ & $N_\tau$  & $T$  [MeV] & $T/T_c$  & $N_{\rm cfg}$  & $N_{\rm src}$ \\
    \hline
  24&128 &  44	 & 0.24 & 171   & 2\\
  24 & 40 & 141 & 0.76 &  301  & 4 \\
  24 & 36 & 156 & 0.84 &  252  & 4 \\
  24 & 32 & 176 & 0.95 & 1000 & 2 \\ 
  24 & 28 & 201 & 1.09 &   501 & 4 \\ 
  24 & 24 & 235 & 1.27 & 1001 & 2 \\  
  24 & 20 & 281 & 1.52 & 1000 & 2 \\ 
  24 & 16 & 352 & 1.90 & 1001 & 2 \\ 
    \hline
   \end{tabular}
\caption{Details of the ensembles. The lattice size is $N_s^3\times N_\tau$, with the temperature $T=1/(a_\tau N_\tau)$. 
  $N_{\rm cfg}$ ($N_{\rm src}$) denotes the number of configurations (sources) used at each volume. 
    The spatial lattice spacing is $a_s=0.1227(8)$ fm, with anisotropy $a_s/a_\tau=3.5$.
  }
\label{tab:lat}
\end{center}
\end{table}


We use our {\sc fastsum} $2+1$ flavour, anisotropic ensembles with the
same parameters as used by the Hadron Spectrum Collaboration
\cite{Edwards:2008ja}. The details of our lattice parameters are
listed in Table \ref{tab:lat}. Our light quarks correspond to
$M_\pi=384(4)$ MeV, with $M_\pi/M_\rho=0.466(3)$;
the strange quark mass is tuned to the physical value
\cite{Lin:2008pr}; and
$T_c$ is determined via the renormalised Polyakov loop.  Details of
these ensembles are available in
Refs.\ \cite{Amato:2013naa,Allton:2014uia,Aarts:2014nba}.

We use Gaussian smearing for the sources and sinks
\cite{Gusken:1989ad} tuned to give the best signal for the positive
parity ground state at the lowest temperature. Our links are
APE-smeared and we use Chroma software \cite{Edwards:2004sx} to
calculate the correlation functions. Details of our procedures are in
\cite{Aarts:2015mma}.



\section{Results}


\begin{figure}
\begin{center}
\includegraphics[width=0.99\textwidth,trim=0 140 70 88,clip]
{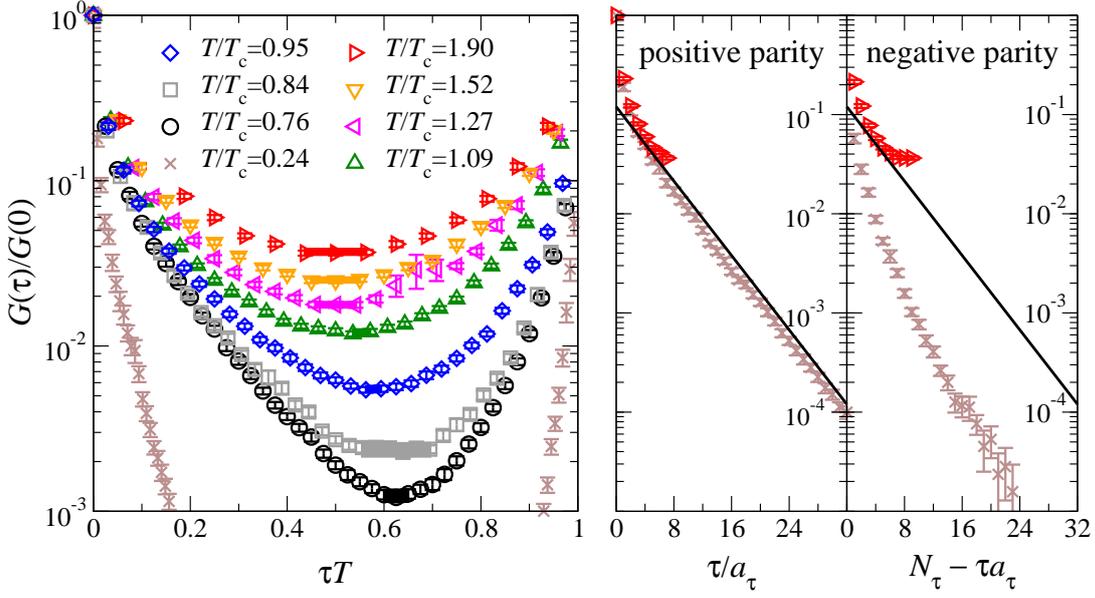}
\caption{{\bf Left pane:} Euclidean correlator $G(\tau)/G(0)$ as a
  function of $\tau T$. At each temperature, the filled rectangle
  indicates where the correlator is a minimum within errors.  {\bf
    Right panes:} The same ratio but versus $\tau/a_\tau$ and for the
  hottest and coldest temperatures only.  The positive and negative
  parity channels are shown separately.  The two (identical) guide
  lines indicate parity non-degeneracy for low $T$ and degeneracy at
  high $T$.
\label{fig:cor}
 }
\end{center}
\end{figure}


The nucleon correlators are shown in the left pane of
Fig.\ \ref{fig:cor} as a function of $\tau T$ for the various
temperatures.  In our case, the $+(-)$ parity states propagate
forwards(backwards) in time.  The data points corresponding to the
minimum of $G(\tau)$ within errors are highlighted with a shaded
rectangle.  As can be seen, for $T\ltap T_c$ this minimum region is
clearly not at the central point, $\tau T = \frac{1}{2}$, whereas for
$T\gtap T_c$ it is.  Therefore, even from direct studies of the correlators
without any fitting procedure, there is evidence that the parity partners
in the nucleon spectrum become degenerate for $T\gtap T_c$.

This analysis is further underlined in right panes of
Fig.\ \ref{fig:cor} where correlators for the hottest and coldest
temperatures are shown separately for the forward ($+$ve parity) and
time-reflected backward ($-$ve parity) parts of the correlators. The
(identical) lines in both graphs are a guide to the eye and again show
clear parity breaking at low temperatures and indicate parity doubling
at high temperatures.

\begin{figure}
\begin{center}
\includegraphics[width=0.99\textwidth,trim=0 140 70 110,clip]
{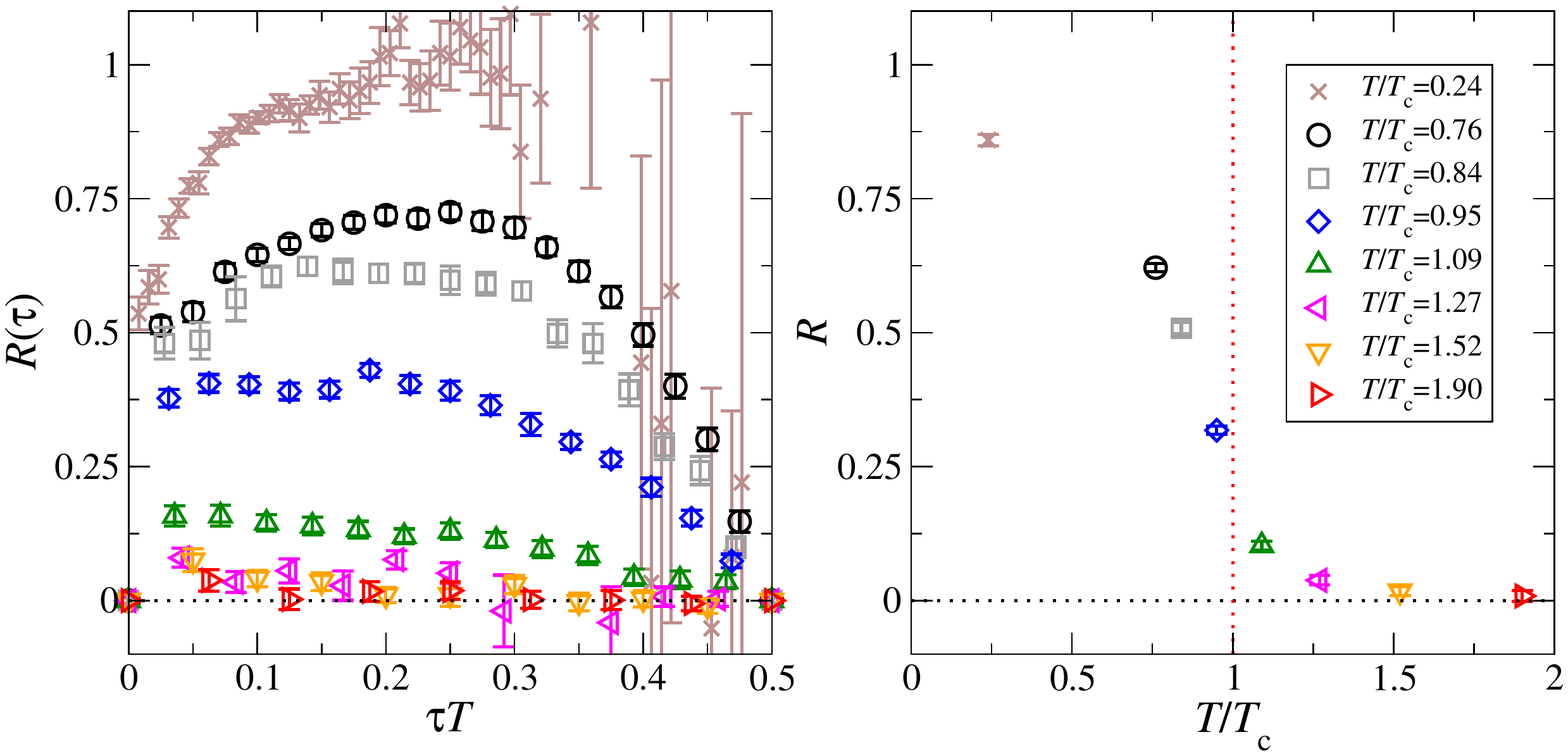}
\caption{
{\bf Left pane:}
The Ratio $R(\tau)$ in Eq.~\protect\ref{eq:Rt} as a function of $\tau T$.
{\bf Right pane:}
Averaged ratio $R$ defined in Eq.~\protect\ref{eq:R} as a function of $T/T_c$.
\label{fig:R}
}
\end{center}
\end{figure}

To study the onset of parity doubling, we define the ratio 
of correlation functions \cite{Datta:2012fz},
\be
\label{eq:Rt}
R(\tau) = \frac{ G(\tau)-G(1/T-\tau)}{ G(\tau) + G(1/T-\tau)}.
\ee
Note that at the central point, $R(\frac{1}{2T})=0$ since
$R(1/T-\tau)=-R(\tau)$. $R(\tau)$ is naturally between 0 and 1: in the
highly non-degenerate case where the $+$ve and $-$ve parity masses
satisfy $M_+\ll M_-$ we have $R(\tau)\approx 1$ (for $\tau\ne
\frac{1}{2T}$), whereas in the degenerate case, we have $R(\tau)\equiv
0$. The results for the ratio are displayed in the left pane of
Fig.\ \ref{fig:R} showing that $R(\tau)\ne 0$ in the hadronic phase,
$T<T_c$, indicating the absence of parity doubling. (The drop towards
zero as $\tau \rightarrow \frac{1}{2T}$ is unavoidable as explained
above.)  As the temperature increases, the ratio decreases and is very
close to zero for $T\gtap T_c$, signally parity doubling.


In order to quantify parity degeneracy further, we consider the
average ratio $R$, defined as
\be
\label{eq:R}
R = \frac{\sum_{n=1}^{\half N_\tau-1} R(\tau_n)/\sigma^2(\tau_n)}
         {\sum_{n=1}^{\half N_\tau-1} 1        /\sigma^2(\tau_n)}, 
\ee
where $\sigma(\tau)$ is the statistical error in $R(\tau)$ and
$\tau_n=na_\tau$.
Again, $R\approx 1$ signals a parity broken phase and $R\approx 0$
a parity-doubling phase.
Figure \ \ref{fig:R} (right pane) plots $R$ showing a clear crossover
behaviour matching our earlier findings of parity doubling at
$T\approx T_c$.

For $T<T_c$, we fit the correlators to standard exponentials obtaining
the masses shown in Table\ \ref{tab:fit} which agree with the results
from \cite{Bulava:2010yg} obtained using variational fitting
techniques.  At the lowest temperature, these masses are higher than
in nature, but this is presumably due to
our quarks being unphysically heavy \cite{Lin:2008pr,Bulava:2010yg}.
From the fit results, $M_+$ has very little temperature dependence
whereas $M_-$ does.  Above $T_c$, exponential fits are found to be
unreliable, presumably due to the ground states no longer being
well-defined particles. We include in the table the dimensionless
ratio,
\be
\Delta=\frac{M_--M_+}{M_-+M_+},
\label{eq:delta}
\ee
to quantify the onset of parity doubling.  In nature, $\Delta=0.241$
at $T=0$, and $\Delta\equiv 0$ in the parity-degenerate phase.  The
table shows that we agree with nature's value at low $T$ (within
errors) and that $\Delta$ then reduces, albeit within large systematic
errors due to the difficulty of performing exponential fits in the
negative parity sector at high temperature.


\begin{table}[t]
\begin{center}
\begin{tabular}{cccccc}
    \hline
      $T/T_c$ & $a_\tau M_+$ & $a_\tau M_-$  & $M_+$ [GeV]  & $M_-$ [GeV]  & $\Delta$\\
    \hline
0.24 & 0.213(5)$\;\;$ & 0.33(5) & 1.20(3)$\;\;$ & 1.9(3) & 0.209(28)(082) \\
0.76 & 0.209(16)  & 0.28(3)   	& 1.18(9)$\;\;$ & 1.6(2) & 0.138(29)(130) \\
0.84 & 0.192(17)  & 0.28(2)   	& 1.08(9)$\;\;$ & 1.6(1) & 0.197(39)(054) \\
0.95 & 0.198(25)  & 0.22(4)   	& 1.12(14) 	& 1.3(2) & 0.052(35)(190) \\
   \hline
   \end{tabular}
\caption{Results from standard exponential fits to the correlators
  below $T_c$ with statistical and systematic errors added in
  quadrature. The final column contains $\Delta$ (see
  Eq.~\protect\ref{eq:delta}) with estimates of statistical and systematic
  errors shown separately. In nature at $T=0$, $\Delta=0.241$.
\label{tab:fit}
}
\end{center}
\end{table}

\begin{figure}
\begin{center}
\includegraphics[width=0.99\textwidth,trim=0 150 20 110,clip]
{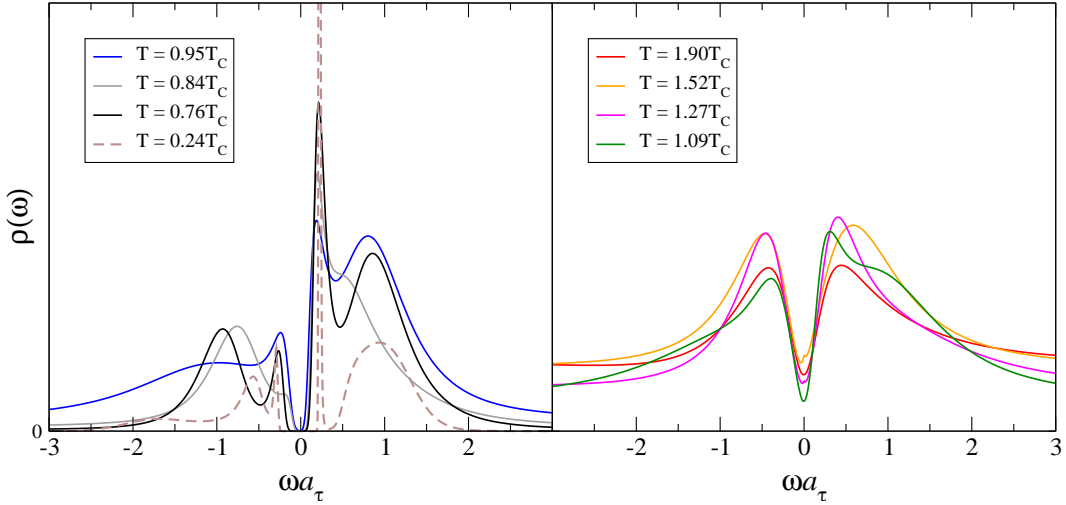}
\caption{The spectral function $\rho(\omega)$ as a function of
  energy, $\omega$, with the left(right) pane showing temperatures
  below(above) $T_c$.  The positive parity channel corresponds to
  $\omega>0$ and the negative parity to $\omega<0$, see
  Eq.~\protect\ref{eq:mem}.
\label{fig:mem}
}
\end{center}
\end{figure}

Further analysis can be made by spectrally decomposing the
correlator $G(\tau)$,
\be
\nn
G(\tau) = \int_0^\infty \frac{d\omega}{2\pi}
\left[ \frac{e^{-\omega      \tau }}{1+e^{-\omega \tau}} \rho_+(\omega)
   -   \frac{e^{-\omega (1/T-\tau)}}{1+e^{-\omega \tau}} \rho_-(\omega)
\right],
\ee
see \cite{kristi:lat15} for details.  We can extract the spectral
functions $\rho_{\pm}(\omega)$ for the $\pm$ve parity states from the
MEM \cite{Asakawa:2000tr} using
an extended integration domain and a modified kernel, $K(\omega,T)$,
\be
\label{eq:mem}
G(\tau) \equiv \int_{-\infty}^{+\infty} \frac{d\omega}{2\pi}
           K(\tau,\omega) \rho(\omega)
\quad\text{and} \quad
\rho_\pm(\omega)
\equiv \rho(\omega)
 \quad\text{for}\quad
 \omega \;\raisebox{-.6ex}{\rlap{$<$}} \raisebox{.5ex}{$>$}\; 0,
\ee
\bea\nn\hspace*{-20mm}
\text{where}\qquad
K(\tau,\omega) &=& \frac{e^{-\omega\tau}}{1+e^{-\omega/T}}
\qquad \;\omega > 0 \\\nn
               &=& \frac{e^{+\omega(1/T-\tau)}}{1+e^{+\omega/T}}
\qquad \omega < 0.  \eea
Details of this procedure will appear shortly \cite{future}, but we
show our main results in Fig.\ \ref{fig:mem} where the spectral
functions are plotted for various temperatures. The MEM analysis
confirms the above picture of a parity non-degenerate phase for
$T\ltap T_c$ with parity doubling, i.e. $\rho_+(\omega) \approx
\rho_-(\omega)$, for $T\gtap T_c$. We also confirm the results for the
ground state masses, $M_\pm$, from the exponential fits (see
Table\ \ref{tab:fit}).


We note that the results shown in this talk used smeared correlation
functions, and thus it is crucial to confirm that no systematic
effects are introduced by this procedure. These checks are performed
in \cite{Aarts:2015mma,future}.



\section*{Conclusions}

We have analysed nucleon correlation functions to determine if parity
doubling occurs for temperatures above $T_c$. We find evidence for
this hypothesis by i) studying the correlators themselves, ii) fitting
the correlators to exponentials, and iii) extracting the positive and
negative parity spectral functions from an MEM analysis.  Since
deconfinement and chiral symmetry restoration are expected to occur
around the same temperature \cite{Borsanyi:2010bp,Bazavov:2011nk}, the
observed parity doubling can be understood from the restoration of
chiral symmetry in the quark-gluon plasma.



\section*{Acknowledgements}

This work used the DiRAC BlueGene/Q Shared Petaflop system at the
University of Edinburgh, operated by the Edinburgh Parallel Computing
Centre on behalf of the STFC DiRAC HPC Facility (www.dirac.ac.uk).
This equipment was funded by BIS National E-infrastructure capital
grant ST/K000411/1, STFC capital grant ST/H008845/1, and STFC DiRAC
Operations grants ST/K005804/1 and ST/K005790/1.
DiRAC is part of the National E-Infrastructure.
We acknowledge the PRACE Grants 2011040469 and Pra05\_1129,
European Union Grant Agreement No. 238353 (ITN STRONGnet),
the STFC grant ST/L000369/1,
and All Souls College Oxford,
HPC Wales,
the Irish Centre for High-End Computing,
the Leverhulme Trust,
the Royal Society,
STFC,
and the Wolfson Foundation
for support.

The authors would like to thank Seyong Kim, Maria Paola Lombardo, Mike
Peardon, Sin\'ead Ryan and Don Sinclair, for useful comments,
discussions and collaboration.





\begin{thebibliography}{99}

\bibitem{Rapp:1999ej}
  R.~Rapp and J.~Wambach,
  Adv.\ Nucl.\ Phys.\  {\bf 25} (2000) 1
  [hep-ph/9909229].

\bibitem{DeTar:1987ar}
  C.~E.~DeTar and J.~B.~Kogut,
  Phys.\ Rev.\ Lett.\  {\bf 59} (1987) 399;
  Phys.\ Rev.\ D {\bf 36} (1987) 2828.

\bibitem{Datta:2012fz}
  S.~Datta, S.~Gupta, M.~Padmanath, J.~Maiti and N.~Mathur,
  JHEP {\bf 1302} (2013) 145
  [arXiv:1212.2927 [hep-lat]].

\bibitem{Pushkina:2004wa}
  I.~Pushkina {\it et al.}  [QCD-TARO Collaboration],
  Phys.\ Lett.\ B {\bf 609} (2005) 265
  [hep-lat/0410017].

\bibitem{Aarts:2015mma}
  G.~Aarts, C.~Allton, S.~Hands, B.~J\"ager, C.~Praki and
  J.~I.~Skullerud,
  Phys.\ Rev.\ D {\bf 92} (2015) 1,  014503
  [arXiv:1502.03603 [hep-lat]].

\bibitem{future}
  G.~Aarts, C.~Allton, S.~Hands, B.~J\"ager, C.~Praki and
  J.~I.~Skullerud,
{\em in preparation}

\bibitem{Gattringer:2010zz}
  C.~Gattringer and C.~B.~Lang,
  ``Quantum chromodynamics on the lattice,''
  Lect.\ Notes Phys.\  {\bf 788} (2010) 1.

\bibitem{Edwards:2008ja}
  R.~G.~Edwards, B.~Jo\'o and H.~W.~Lin,
  Phys.\ Rev.\ D {\bf 78} (2008) 054501
  [arXiv:0803.3960 [hep-lat]].

\bibitem{Lin:2008pr}
  H.~W.~Lin {\it et al.}  [Hadron Spectrum Collaboration],
  Phys.\ Rev.\ D {\bf 79} (2009) 034502
  [arXiv:0810.3588 [hep-lat]].

\bibitem{Amato:2013naa}
  A.~Amato, G.~Aarts, C.~Allton, P.~Giudice, S.~Hands and J.~I.~Skullerud,
  Phys.\ Rev.\ Lett.\  {\bf 111} (2013) 172001
  [arXiv:1307.6763 [hep-lat]].

\bibitem{Allton:2014uia}
  C.~Allton {\it et al.},
  PoS LATTICE {\bf 2013} (2014) 151
  [arXiv:1401.2116 [hep-lat]].

\bibitem{Aarts:2014nba}
  G.~Aarts, C.~Allton, A.~Amato, P.~Giudice, S.~Hands and J.~I.~Skullerud,
  JHEP {\bf 1502} (2015) 186
  [arXiv:1412.6411 [hep-lat]].

\bibitem{Gusken:1989ad}
  S.~G\"usken, U.~Low, K.~H.~M\"utter, R.~Sommer, A.~Patel and K.~Schilling,
  Phys.\ Lett.\ B {\bf 227} (1989) 266.
 
\bibitem{Edwards:2004sx}
  R.~G.~Edwards {\it et al.}  [SciDAC and LHPC and UKQCD Collaborations],
  Nucl.\ Phys.\ Proc.\ Suppl.\  {\bf 140} (2005) 832
  [hep-lat/0409003].
 
\bibitem{Bulava:2010yg}
  J.~Bulava {\it et al}.,
  Phys.\ Rev.\ D {\bf 82} (2010) 014507
  [arXiv:1004.5072 [hep-lat]].

\bibitem{kristi:lat15}
  C.~Praki and G.~Aarts,
  Talk presented at Lattice 2015, Kobe, Japan, July 2015,
  PoS(LATTICE 2015)182

\bibitem{Asakawa:2000tr}
  M.~Asakawa, T.~Hatsuda and Y.~Nakahara,
  Prog.\ Part.\ Nucl.\ Phys.\  {\bf 46} (2001) 459
  [hep-lat/0011040].

\bibitem{Borsanyi:2010bp}
  S.~Bors\'anyi {\it et al.}  [Wuppertal-Budapest Collaboration],
  JHEP {\bf 1009} (2010) 073
  [arXiv:1005.3508 [hep-lat]].

\bibitem{Bazavov:2011nk}
  A.~Bazavov, T.~Bhattacharya, M.~Cheng, C.~DeTar, H.~T.~Ding, S.~Gottlieb and  R.~Gupta {\it et al.},
  Phys.\ Rev.\ D {\bf 85} (2012) 054503
  [arXiv:1111.1710 [hep-lat]].

\end{thebibliography}
\end{document}